\documentstyle[12pt]{article}
\def\beq{\begin{equation}}
\def\barr{\begin{eqnarray}}
\def\ra{\rightarrow}
\def\eeq{\end{equation}}
\def\err{\end{eqnarray}}

\def\bea{\begin{eqnarray}}
\def\eea{\end{eqnarray}}

\begin{document}

\begin{flushright}

hep-ph/9507402 \\
HD-THEP-95-28 \\
IOA-226-95 \\
\end{flushright}
\vspace{15mm}

\begin{center}
{\large {\bf MSSM phenomenology in the large
\mbox{\boldmath{Tan$\beta$}} regime}}
\end{center}

\vspace*{0.4cm}

\begin{center}
{\bf E. G. Floratos $^{1,*}$,
G. K. Leontaris $^{2,**}$} and {\bf S. Lola$^{3}$}
\end{center}

\begin{center}

\begin{tabular}{c}
$^{1}$ {\small Institute of Nuclear Physics, NSRC  Demokritos,}\\
{\small  Athens, Greece}\\

$^{2}${\small
Centre de Physique Th\'eorique, Ecole Polytechnique},\\
{\small F-91128 Palaiseau, France} \\

$^{3}${\small Institut f\"{u}r
Theoretische Physik, Univerisit\"at
Heidelberg,}\\
{\small Philosophenweg 16, 69120 Heidelberg, Germany }\\
\end{tabular}
\end{center}

\vspace{1 cm}

\begin{center}
{\bf ABSTRACT}
\end{center}

\noindent
 We discuss aspects of the low energy phenomenology of the
MSSM, in  the large  $\tan {\beta} $  regime. We
explore the regions of the parameter space
where the  $h_t$ and $h_b$ Yukawa
couplings exhibit a fixed point structure, using previous
analytic solutions for these couplings.
Expressions for the parameters $A_{t}$ and $A_{b}$  and the
renormalised soft mass terms are also derived, making it possible
to estimate analytically the  sparticle loop -- corrections
to the  bottom mass, which are important in this limit.

\thispagestyle{empty}
\setcounter{page}{0}
\vfill

\noindent July 1995

\vspace{.5cm}
\hrule
\vspace{.3cm}
{\small
\noindent
{\small
$^{(*)}$On leave of absence from Phys. Dept., University of Crete,
Iraklion,Crete, Greece.} \\
{\small $^{(**)}$On leave of absence from Phys. Dept.,
University of Ioannina,
451 10 Ioannina, Greece.}

\newpage

\section{Introduction}

Among the extensions of the Standard Model (SM), supersymmetry
seems to provide the best grounds towards a unification
 of the fundamental interactions.
This is not only because of the natural solution to the hierarchy
problem \cite{susy},
but also because of the correct prediction of
$\sin^{2}{\theta_{W}}$, and the convergence of the three gauge
couplings at a point, at a scale $O(10^{16})$ GeV, unlike in the
non--supersymmetric grand unified schemes \cite{unif}.
The simplest supersymmetric extension of the theory is the
Minimal Supersymmetric Standard Model (MSSM) which has the minimal
number of fields and Yukawa couplings that is consistent with the
content of the SM. However, even in the MSSM
several arbitrary parameters also exist.
These are the initial conditions  for the Yukawa
couplings and the scalar masses, which are expected to be
fixed at an even more fundamental level, like string theory.
A simplified approach to reduce the number of these variables,
would be to assume universality of all scalar masses
at the unification scale. Then, one is left with only five new
arbitrary parameters in addition to those of the non--supersymmetric
standard model.  However, while this is consistent
with the low $\tan{\beta}$ regime, in the large $\tan{\beta}$
limit of the theory (where $h_{t} \approx h_{b}$),
in order to get the correct radiative electroweak symmetry breaking
pattern \cite{radi} one is forced to depart from universality.
This latter case, i.e. large tan$\beta$ and non-universal boundary
conditions for the scalars,  appears naturally in many string derived
models. In fact it is generally expected that irrespectively of the
fundamental theory, the low energy model will look much like the
supersymmetric standard model with non--universal boundary conditions
for the soft terms. Further theoretical expectations suggest that
the Yukawa couplings of the third generation are of the order of
the  common gauge coupling at the unification scale. Moreover,
specific grand unified groups suggest equalities of the form
$h_t = h_b = h_{\tau}$ at $M_{U}$.

It is interesting to investigate whether radiative
corrections  can  determine the Yukawa couplings and possibly the
soft masses  of the effective low energy  theory.
For example, the Yukawa couplings  are running quantities
from the unification scale $M_U$ down  to low energy.
If they are relatively large at $M_U$,
their low energy values exhibit a fixed point structure\cite{ross}
and they are rather insensitive to their initial conditions.
The analysis of the above becomes more interesting by the fact that
there is recent experimental evidence for a top quark with
a mass of ${\cal O}(180GeV)$ which is compatible with the
prediction of the fixed point structure.

In the present letter we investigate the fixed point structure
of previous analytic solutions for the $h_t,h_b$ couplings
\cite{fl,kpz1,kpz2}.
We further extend the existing analytic solutions \cite{fl1}
for the scalar masses in the case of $tan\beta \gg 1$,
by including the contribution from the
$A_t$ and $A_b$ terms. This is
of particular importance not only for the determination of
the scalar mass spectrum itself, but for the correct theoretical
computation of the fermion  masses. In particular, when $tan\beta\gg1$
the bottom mass receives large corrections from sparticle loops
\cite{bl,sc}.
The tau lepton receives also corrections of the same type but
they are less significant. A precise determination of these
corrections would require the knowledge of the
scalar masses  involved in the computation.
An analytic approach in both Yukawa and soft mass terms would
make possible a systematic exploration of these corrections.

\section{Yukawa coupling fixed point}

In this section we are going to use the results of
\cite{fl1} for the top and bottom
Yukawa couplings, in order to identify the regions where these
couplings exhibit a fixed point behaviour and give simplified expressions
that describe these fixed points.
The renormalisation group equations for the top-bottom
system (when ignoring the $h_{\tau}$ Yukawa coupling), read
\begin{eqnarray}
\frac{d}{dt} h^2_t &=&
\frac{1}{8\pi^2}
 \Big\{6h^2_t + h^2_b - G_Q \Big\} h^2_t \label{eq:1} \\
\frac{d}{dt} h^2_b &=& \frac{1}{8\pi^2}
\Big\{h^2_t + 6h^2_b - G_D \Big\} h^2_b \label{eq:2}
\end{eqnarray}
where
\begin{equation}
G_Q = \sum^{3}_{i=1} c^i_Q g^2_i\,\, ,\qquad  G_D = \sum^{3}_{i=1}
 c^i_D g^2_i
\label{eq:3}
\end{equation}
Here  $t = \ln{Q}$, where $Q$ is the energy scale,
$c^i_Q = \Big\{
\frac{13}{15}, 3, \frac{16}{3}\Big\}$ and
$ c^i_B = \Big\{\frac{7}{15}, 3,
\frac{16}{3}\Big\}$ for
$U(1)$, $SU(2)$ and $SU(3)$
 respectively.
Ignoring small $U(1)$ corrections, this
system can be solved.
Defining the parameters $x,y$ through
 $h^2_t = \gamma^2_Q x ,\,\,   h^2_b =
\gamma^2_Q y $,
where
\beq
\gamma^2_Q =  \exp \Big\{\frac{1}
{8\pi^2} \int_{t}^{t_0}  G_Q({t^{\prime}})
dt^{\prime}\Big\} \equiv
\prod_{j=1}^3
\left(\frac{\alpha_{j,0}}{\alpha_{j}}\right)^{\frac{c_Q^j}{b_j}}
\label{eq:gammaQ}
\eeq
 one can make the transformation
\begin{equation}
u=\frac{k_0}{(x-y)^{5/6}}\equiv \frac{k_0}{{\omega}^{5/6}},
\qquad d\,I=\frac{6}{8\pi^2}\gamma_Q^2 d\,t \label{uvar}
\end{equation}
where $\omega = x - y$
and  $I(t) =\frac{3}
{4\pi^2} \int^t_{t_0} \gamma_Q^2({t^{\prime}})
dt^{\prime}$.
The parameter
\beq
k_0 = 4 \frac{x_0y_0}{(x_0-y_0)^{7/6}}
\eeq
depends on the
initial conditions $x_0\equiv h_{t,0}^2$ and $y_0\equiv h_{b,0}^2$.
Then, one forms a differential equation for the new  variable $u$,
which can be solved in terms of hypergeometric functions. In particular,
the $t-b$ Yukawa coupling solutions can be expressed in
terms of the variable $u$ as follows\cite{fl} :
\begin{eqnarray}
h_t^2 &\equiv & \frac{1}{2}\gamma^2_Q \omega (\sqrt{1+u}+1)
\label{ht}\\
h_b^2&\equiv &  \frac{1}{2}\gamma^2_D \omega (\sqrt{1+u}-1)
\label{hb}
\end{eqnarray}
Note that we have ``restored'' the symmetry between the
differential equations (\ref{eq:1}-\ref{eq:2}) and the solutions
(\ref{ht}-\ref{hb}) by replacing  the gauge factor  $\gamma^2_Q$
with $\gamma^2_D$ in  (\ref{hb}).

It is interesting to search for particular combinations of
the Yukawa couplings which are rather independent of
their initial values. To start our investigation we first obtain
a simplified formula for the function $\omega (t)= x - y$,
in the case $h_{b,t}\ge 1$
\beq
\omega (t) \approx \frac{x_0-y_0}{\{_2F_1^0 + \frac 76
\sqrt{x_0y_0}I(t)\}^{12/7}}
\label{wm}
\eeq
where $_2F_1^0$  is the value of the hypergeometric function at $u=u_0$.
 Then, we can use  the  relation between $x,y$  variables\cite{fl}
\begin{equation}
\Big(\frac {x-y} {x_0-y_0}\Big)^7 = \Big(\frac{xy}
{x_0y_0}\Big)^6 .  \label{eq:xy}
\end{equation}
to obtain the following expression,
\beq
\sqrt{x y}= \frac{\sqrt{x_0 y_0}}{_2F_1^0 + \frac 76
\sqrt{x_0y_0}I(t)}\label{xy}
\eeq
The above equation can also easily be obtained by a direct
multiplication of the solutions (\ref{ht},\ref{hb}) substituting
$u$ from (\ref{uvar}).

Note first that for a small
difference between $h_{t}$ and $h_{b}$,
${}_2F_1^0 \approx 1$, while  the integral $I\ge 10$.
Thus for  $\sqrt{x_0 y_0}\equiv h_{t,0}h_{b,0}\ge 1$ we can write
\beq
h_t h_b \approx
\frac{{8\pi^2}\gamma_Q \gamma_D}{{7}\int \gamma_Q^2 d\,t}
\label{fxd}
\eeq
This last expression tells us that we can get an approximate,
model independent  prediction for the product of $h_t,h_b$ couplings
at the low energy scale provided we start with relatively large
and comparable $h_{t,0},h_{b,0}$ values  at $M_U$.
We note in passing, that
it is possible to use Yukawa coupling constraints obtained
at the Unification scale to eliminate one of the two parameters.
Suitable constraints combined with (\ref{fxd}), can determine the
absolute values of both couplings. In \cite{BD} for example,
it is shown that within spontaneously broken $N=1$ supergravity
models two generic types of  such constraints are obtained.  For a
superpotential ${\cal W}=h_{i_Ai_Bi_C}\Phi_{i_A}\Phi_{i_B}\Phi_{i_C}$
while assuming different scale structures for the various
fields in the K\"ahler potential, there are  multiplicative
(duality invariant) constraints of the form $\prod_{i_Ai_Bi_C}
h_{i_Ai_Bi_C} = cst$ if of course  $h_{i_Ai_Bi_C} \not= 0$.
Even if this constraint were applicable for the two Yukawa couplings
which interest us here,  it could not be useful  however, due to the
fixed point property of the product in relation (\ref{fxd}).
A more interesting situation  arises in cases where the constraint
applies to the ratio of the Yukawa couplings.  A simple example is
shown in \cite{BD} for  two couplings  with a K\"ahler potential
having the two fields in the same no--scale structure. Thus if the
higher theory can give a prediction about the ratio of the two
Yukawas  at $M_U$, we can use this in conjunction with the analytic
solution to extract information about the  Yukawas at the weak scale.

Then the low energy values can be obtained using the relation
 (\ref{fxd}) and the low energy ratio obtained from the
analytic solution in the limit $h_{t,0}h_{b,0} >1$,
\beq
\frac{h_t^2}{h_b^2}\approx \frac{\gamma_Q^2}{\gamma_D^2}
\frac{\sqrt{1+u}+1}{\sqrt{1+u}-1}\label{fxd1}
\eeq
For the limit of interest,  $u$ is given by the  approximate
formula
\beq
u\approx u_0 \{1 + \frac{7}{6}p_0I(t)\}^{10/7}
\eeq
with
\beq
u_0 = \left({2 r_0}/({r_0^2-1})\right)^2
\eeq
and $r_0,p_0$ the ratio and product of $h_{t,0},h_{b,0}$
couplings respectively.
Thus in the case of $r_0\ra 1$, $u\ra \infty$, and the ratio of
the Yukawas runs also to a fixed point value determined
approximately by the ratio $\gamma_Q/\gamma_D$.

In Figs. 1a, 1b and 1c, we show the fixed point structure of the
Yukawa couplings and their product, when we use the
analytic expressions without making
any approximation on the hyper-geometric
function\footnote{The fact that we use
expressions up to only one-loop, as well as the approximations
made in order to derive the analytic formulas for
the couplings, cause small errors in the numerical
values of
$h_{t}$ and $h_{b}$ which however do
not alter the validity of the results.}.
We have taken a common gauge coupling
$a_G = 1/25.0$ at a unification scale
$m_G = 1.35 \cdot 10^{16}$ GeV and a supersymmetry
breaking scale $\approx 200$ GeV, leading to an
$a_{s}(M_{z}) \approx 0.112$. The couplings are
presented at the top mass.

Fig. 1a shows $h_t$-low energy coupling  versus the
GUT $h_t^0$ values for the ratios $h_t^0/h_b^0 =1.2$
and $1.001$ denoted with stars and crosses respectively.
 For a wide $h_t^0$ range, the estimated
$h_t(m_t)$ values differ at most by $1\%$.

 Fig. 1b shows  the bottom coupling versus $h_b^0$.
Here we took the same region of initial values for
$h_t^0$ for each of the two cases quoted,
thus the two lines are interrupted before touching
the contour.
In Fig. 1c we plot  the product $h_b h_t$
for the same input ratios as in fig 1a.
In Fig. 1c, a rather interesting fixed point property
is exhibited if $h_t^0h_b^0\ge 4$ were the  estimated
low energy values  differ in less than  $1\%$.
On the other hand, it is remarkable that
different initial $h_t^0/h_b^0$ ratios
accumulate exactly on the same curve at the weak scale.
This is in a very good agreement with the equations
that we derived for the description of the fixed points
in terms of functions of the gauge couplings only.

\setlength{\unitlength}{0.240900pt}
\ifx\plotpoint\undefined\newsavebox{\plotpoint}\fi
\sbox{\plotpoint}{\rule[-0.200pt]{0.400pt}{0.400pt}}%
\begin{picture}(1500,629)(0,0)
\font\gnuplot=cmr10 at 10pt
\gnuplot
\sbox{\plotpoint}{\rule[-0.200pt]{0.400pt}{0.400pt}}%
\put(220.0,143.0){\rule[-0.200pt]{4.818pt}{0.400pt}}
\put(198,143){\makebox(0,0)[r]{1.6}}
\put(1416.0,143.0){\rule[-0.200pt]{4.818pt}{0.400pt}}
\put(220.0,203.0){\rule[-0.200pt]{4.818pt}{0.400pt}}
\put(198,203){\makebox(0,0)[r]{1.8}}
\put(1416.0,203.0){\rule[-0.200pt]{4.818pt}{0.400pt}}
\put(220.0,262.0){\rule[-0.200pt]{4.818pt}{0.400pt}}
\put(198,262){\makebox(0,0)[r]{2}}
\put(1416.0,262.0){\rule[-0.200pt]{4.818pt}{0.400pt}}
\put(220.0,322.0){\rule[-0.200pt]{4.818pt}{0.400pt}}
\put(198,322){\makebox(0,0)[r]{2.2}}
\put(1416.0,322.0){\rule[-0.200pt]{4.818pt}{0.400pt}}
\put(220.0,382.0){\rule[-0.200pt]{4.818pt}{0.400pt}}
\put(198,382){\makebox(0,0)[r]{2.4}}
\put(1416.0,382.0){\rule[-0.200pt]{4.818pt}{0.400pt}}
\put(220.0,442.0){\rule[-0.200pt]{4.818pt}{0.400pt}}
\put(198,442){\makebox(0,0)[r]{2.6}}
\put(1416.0,442.0){\rule[-0.200pt]{4.818pt}{0.400pt}}
\put(220.0,501.0){\rule[-0.200pt]{4.818pt}{0.400pt}}
\put(198,501){\makebox(0,0)[r]{2.8}}
\put(1416.0,501.0){\rule[-0.200pt]{4.818pt}{0.400pt}}
\put(220.0,561.0){\rule[-0.200pt]{4.818pt}{0.400pt}}
\put(198,561){\makebox(0,0)[r]{3}}
\put(1416.0,561.0){\rule[-0.200pt]{4.818pt}{0.400pt}}
\put(342.0,113.0){\rule[-0.200pt]{0.400pt}{4.818pt}}
\put(342,68){\makebox(0,0){0.96}}
\put(342.0,541.0){\rule[-0.200pt]{0.400pt}{4.818pt}}
\put(585.0,113.0){\rule[-0.200pt]{0.400pt}{4.818pt}}
\put(585,68){\makebox(0,0){0.98}}
\put(585.0,541.0){\rule[-0.200pt]{0.400pt}{4.818pt}}
\put(828.0,113.0){\rule[-0.200pt]{0.400pt}{4.818pt}}
\put(828,68){\makebox(0,0){1}}
\put(828.0,541.0){\rule[-0.200pt]{0.400pt}{4.818pt}}
\put(1071.0,113.0){\rule[-0.200pt]{0.400pt}{4.818pt}}
\put(1071,68){\makebox(0,0){1.02}}
\put(1071.0,541.0){\rule[-0.200pt]{0.400pt}{4.818pt}}
\put(1314.0,113.0){\rule[-0.200pt]{0.400pt}{4.818pt}}
\put(1314,68){\makebox(0,0){1.04}}
\put(1314.0,541.0){\rule[-0.200pt]{0.400pt}{4.818pt}}
\put(220.0,113.0){\rule[-0.200pt]{292.934pt}{0.400pt}}
\put(1436.0,113.0){\rule[-0.200pt]{0.400pt}{107.923pt}}
\put(220.0,561.0){\rule[-0.200pt]{292.934pt}{0.400pt}}
\put(45,337){\makebox(0,0){\mbox{\boldmath{$h_{t}^0$}}}}
\put(828,23){\makebox(0,0){\mbox{\boldmath{$h_{t}$}}}}
\put(828,606){\makebox(0,0){{ $h_t$ - fixed point structure for
 \mbox{\boldmath{$h_{t}^0/h_{b}^0=1.2,1.001$}}}}}
\put(220.0,113.0){\rule[-0.200pt]{0.400pt}{107.923pt}}
\put(799,113){\makebox(0,0){$+$}}
\put(815,128){\makebox(0,0){$+$}}
\put(830,143){\makebox(0,0){$+$}}
\put(844,158){\makebox(0,0){$+$}}
\put(856,173){\makebox(0,0){$+$}}
\put(867,188){\makebox(0,0){$+$}}
\put(878,203){\makebox(0,0){$+$}}
\put(887,218){\makebox(0,0){$+$}}
\put(896,232){\makebox(0,0){$+$}}
\put(905,247){\makebox(0,0){$+$}}
\put(912,262){\makebox(0,0){$+$}}
\put(919,277){\makebox(0,0){$+$}}
\put(926,292){\makebox(0,0){$+$}}
\put(932,307){\makebox(0,0){$+$}}
\put(938,322){\makebox(0,0){$+$}}
\put(943,337){\makebox(0,0){$+$}}
\put(948,352){\makebox(0,0){$+$}}
\put(953,367){\makebox(0,0){$+$}}
\put(958,382){\makebox(0,0){$+$}}
\put(962,397){\makebox(0,0){$+$}}
\put(966,412){\makebox(0,0){$+$}}
\put(969,427){\makebox(0,0){$+$}}
\put(973,442){\makebox(0,0){$+$}}
\put(976,456){\makebox(0,0){$+$}}
\put(979,471){\makebox(0,0){$+$}}
\put(982,486){\makebox(0,0){$+$}}
\put(985,501){\makebox(0,0){$+$}}
\put(988,516){\makebox(0,0){$+$}}
\put(990,531){\makebox(0,0){$+$}}
\put(993,546){\makebox(0,0){$+$}}
\put(995,561){\makebox(0,0){$+$}}
\put(870,113){\makebox(0,0){$\star$}}
\put(884,128){\makebox(0,0){$\star$}}
\put(897,143){\makebox(0,0){$\star$}}
\put(909,158){\makebox(0,0){$\star$}}
\put(919,173){\makebox(0,0){$\star$}}
\put(929,188){\makebox(0,0){$\star$}}
\put(938,203){\makebox(0,0){$\star$}}
\put(946,218){\makebox(0,0){$\star$}}
\put(953,232){\makebox(0,0){$\star$}}
\put(960,247){\makebox(0,0){$\star$}}
\put(966,262){\makebox(0,0){$\star$}}
\put(972,277){\makebox(0,0){$\star$}}
\put(977,292){\makebox(0,0){$\star$}}
\put(982,307){\makebox(0,0){$\star$}}
\put(987,322){\makebox(0,0){$\star$}}
\put(991,337){\makebox(0,0){$\star$}}
\put(995,352){\makebox(0,0){$\star$}}
\put(999,367){\makebox(0,0){$\star$}}
\put(1002,382){\makebox(0,0){$\star$}}
\put(1005,397){\makebox(0,0){$\star$}}
\put(1008,412){\makebox(0,0){$\star$}}
\put(1011,427){\makebox(0,0){$\star$}}
\put(1014,442){\makebox(0,0){$\star$}}
\put(1016,456){\makebox(0,0){$\star$}}
\put(1018,471){\makebox(0,0){$\star$}}
\put(1021,486){\makebox(0,0){$\star$}}
\put(1023,501){\makebox(0,0){$\star$}}
\put(1025,516){\makebox(0,0){$\star$}}
\put(1026,531){\makebox(0,0){$\star$}}
\put(1028,546){\makebox(0,0){$\star$}}
\put(1030,561){\makebox(0,0){$\star$}}
\end{picture}
\begin{center}
{\bf Fig. 1a}
\end{center}

\vspace{1.3 cm}

\setlength{\unitlength}{0.240900pt}
\ifx\plotpoint\undefined\newsavebox{\plotpoint}\fi
\sbox{\plotpoint}{\rule[-0.200pt]{0.400pt}{0.400pt}}%
\begin{picture}(1500,629)(0,0)
\font\gnuplot=cmr10 at 10pt
\gnuplot
\sbox{\plotpoint}{\rule[-0.200pt]{0.400pt}{0.400pt}}%
\put(220.0,113.0){\rule[-0.200pt]{4.818pt}{0.400pt}}
\put(198,113){\makebox(0,0)[r]{1.2}}
\put(1416.0,113.0){\rule[-0.200pt]{4.818pt}{0.400pt}}
\put(220.0,163.0){\rule[-0.200pt]{4.818pt}{0.400pt}}
\put(198,163){\makebox(0,0)[r]{1.4}}
\put(1416.0,163.0){\rule[-0.200pt]{4.818pt}{0.400pt}}
\put(220.0,213.0){\rule[-0.200pt]{4.818pt}{0.400pt}}
\put(198,213){\makebox(0,0)[r]{1.6}}
\put(1416.0,213.0){\rule[-0.200pt]{4.818pt}{0.400pt}}
\put(220.0,262.0){\rule[-0.200pt]{4.818pt}{0.400pt}}
\put(198,262){\makebox(0,0)[r]{1.8}}
\put(1416.0,262.0){\rule[-0.200pt]{4.818pt}{0.400pt}}
\put(220.0,312.0){\rule[-0.200pt]{4.818pt}{0.400pt}}
\put(198,312){\makebox(0,0)[r]{2}}
\put(1416.0,312.0){\rule[-0.200pt]{4.818pt}{0.400pt}}
\put(220.0,362.0){\rule[-0.200pt]{4.818pt}{0.400pt}}
\put(198,362){\makebox(0,0)[r]{2.2}}
\put(1416.0,362.0){\rule[-0.200pt]{4.818pt}{0.400pt}}
\put(220.0,412.0){\rule[-0.200pt]{4.818pt}{0.400pt}}
\put(198,412){\makebox(0,0)[r]{2.4}}
\put(1416.0,412.0){\rule[-0.200pt]{4.818pt}{0.400pt}}
\put(220.0,461.0){\rule[-0.200pt]{4.818pt}{0.400pt}}
\put(198,461){\makebox(0,0)[r]{2.6}}
\put(1416.0,461.0){\rule[-0.200pt]{4.818pt}{0.400pt}}
\put(220.0,511.0){\rule[-0.200pt]{4.818pt}{0.400pt}}
\put(198,511){\makebox(0,0)[r]{2.8}}
\put(1416.0,511.0){\rule[-0.200pt]{4.818pt}{0.400pt}}
\put(220.0,561.0){\rule[-0.200pt]{4.818pt}{0.400pt}}
\put(198,561){\makebox(0,0)[r]{3}}
\put(1416.0,561.0){\rule[-0.200pt]{4.818pt}{0.400pt}}
\put(220.0,113.0){\rule[-0.200pt]{0.400pt}{4.818pt}}
\put(220,68){\makebox(0,0){0.92}}
\put(220.0,541.0){\rule[-0.200pt]{0.400pt}{4.818pt}}
\put(342.0,113.0){\rule[-0.200pt]{0.400pt}{4.818pt}}
\put(342,68){\makebox(0,0){0.93}}
\put(342.0,541.0){\rule[-0.200pt]{0.400pt}{4.818pt}}
\put(463.0,113.0){\rule[-0.200pt]{0.400pt}{4.818pt}}
\put(463,68){\makebox(0,0){0.94}}
\put(463.0,541.0){\rule[-0.200pt]{0.400pt}{4.818pt}}
\put(585.0,113.0){\rule[-0.200pt]{0.400pt}{4.818pt}}
\put(585,68){\makebox(0,0){0.95}}
\put(585.0,541.0){\rule[-0.200pt]{0.400pt}{4.818pt}}
\put(706.0,113.0){\rule[-0.200pt]{0.400pt}{4.818pt}}
\put(706,68){\makebox(0,0){0.96}}
\put(706.0,541.0){\rule[-0.200pt]{0.400pt}{4.818pt}}
\put(828.0,113.0){\rule[-0.200pt]{0.400pt}{4.818pt}}
\put(828,68){\makebox(0,0){0.97}}
\put(828.0,541.0){\rule[-0.200pt]{0.400pt}{4.818pt}}
\put(950.0,113.0){\rule[-0.200pt]{0.400pt}{4.818pt}}
\put(950,68){\makebox(0,0){0.98}}
\put(950.0,541.0){\rule[-0.200pt]{0.400pt}{4.818pt}}
\put(1071.0,113.0){\rule[-0.200pt]{0.400pt}{4.818pt}}
\put(1071,68){\makebox(0,0){0.99}}
\put(1071.0,541.0){\rule[-0.200pt]{0.400pt}{4.818pt}}
\put(1193.0,113.0){\rule[-0.200pt]{0.400pt}{4.818pt}}
\put(1193,68){\makebox(0,0){1}}
\put(1193.0,541.0){\rule[-0.200pt]{0.400pt}{4.818pt}}
\put(1314.0,113.0){\rule[-0.200pt]{0.400pt}{4.818pt}}
\put(1314,68){\makebox(0,0){1.01}}
\put(1314.0,541.0){\rule[-0.200pt]{0.400pt}{4.818pt}}
\put(1436.0,113.0){\rule[-0.200pt]{0.400pt}{4.818pt}}
\put(1436,68){\makebox(0,0){1.02}}
\put(1436.0,541.0){\rule[-0.200pt]{0.400pt}{4.818pt}}
\put(220.0,113.0){\rule[-0.200pt]{292.934pt}{0.400pt}}
\put(1436.0,113.0){\rule[-0.200pt]{0.400pt}{107.923pt}}
\put(220.0,561.0){\rule[-0.200pt]{292.934pt}{0.400pt}}
\put(45,337){\makebox(0,0){\mbox{\boldmath{$h_{b}^0$}}}}
\put(828,23){\makebox(0,0){\mbox{\boldmath{$h_{b}$}}}}
\put(828,606){\makebox(0,0){{ $h_b$ - fixed point structure
 for \mbox{\boldmath{$h_{t}^0/h_{b}^0=1.2,1.001$}}}}}
\put(220.0,113.0){\rule[-0.200pt]{0.400pt}{107.923pt}}
\put(854,187){\makebox(0,0){$+$}}
\put(870,200){\makebox(0,0){$+$}}
\put(884,212){\makebox(0,0){$+$}}
\put(897,225){\makebox(0,0){$+$}}
\put(909,237){\makebox(0,0){$+$}}
\put(921,249){\makebox(0,0){$+$}}
\put(931,262){\makebox(0,0){$+$}}
\put(940,274){\makebox(0,0){$+$}}
\put(949,287){\makebox(0,0){$+$}}
\put(957,299){\makebox(0,0){$+$}}
\put(965,312){\makebox(0,0){$+$}}
\put(971,324){\makebox(0,0){$+$}}
\put(978,336){\makebox(0,0){$+$}}
\put(984,349){\makebox(0,0){$+$}}
\put(990,361){\makebox(0,0){$+$}}
\put(995,374){\makebox(0,0){$+$}}
\put(1000,386){\makebox(0,0){$+$}}
\put(1005,399){\makebox(0,0){$+$}}
\put(1009,411){\makebox(0,0){$+$}}
\put(1013,424){\makebox(0,0){$+$}}
\put(1017,436){\makebox(0,0){$+$}}
\put(1020,448){\makebox(0,0){$+$}}
\put(1024,461){\makebox(0,0){$+$}}
\put(1027,473){\makebox(0,0){$+$}}
\put(1030,486){\makebox(0,0){$+$}}
\put(1033,498){\makebox(0,0){$+$}}
\put(1036,511){\makebox(0,0){$+$}}
\put(1039,523){\makebox(0,0){$+$}}
\put(1041,535){\makebox(0,0){$+$}}
\put(1043,548){\makebox(0,0){$+$}}
\put(1046,560){\makebox(0,0){$+$}}
\put(671,125){\makebox(0,0){$\star$}}
\put(695,136){\makebox(0,0){$\star$}}
\put(717,146){\makebox(0,0){$\star$}}
\put(738,157){\makebox(0,0){$\star$}}
\put(757,167){\makebox(0,0){$\star$}}
\put(775,177){\makebox(0,0){$\star$}}
\put(791,188){\makebox(0,0){$\star$}}
\put(806,198){\makebox(0,0){$\star$}}
\put(820,208){\makebox(0,0){$\star$}}
\put(833,219){\makebox(0,0){$\star$}}
\put(845,229){\makebox(0,0){$\star$}}
\put(856,240){\makebox(0,0){$\star$}}
\put(867,250){\makebox(0,0){$\star$}}
\put(877,260){\makebox(0,0){$\star$}}
\put(886,271){\makebox(0,0){$\star$}}
\put(895,281){\makebox(0,0){$\star$}}
\put(903,291){\makebox(0,0){$\star$}}
\put(911,302){\makebox(0,0){$\star$}}
\put(918,312){\makebox(0,0){$\star$}}
\put(925,322){\makebox(0,0){$\star$}}
\put(932,333){\makebox(0,0){$\star$}}
\put(938,343){\makebox(0,0){$\star$}}
\put(944,354){\makebox(0,0){$\star$}}
\put(949,364){\makebox(0,0){$\star$}}
\put(954,374){\makebox(0,0){$\star$}}
\put(959,385){\makebox(0,0){$\star$}}
\put(964,395){\makebox(0,0){$\star$}}
\put(969,405){\makebox(0,0){$\star$}}
\put(973,416){\makebox(0,0){$\star$}}
\put(977,426){\makebox(0,0){$\star$}}
\put(981,437){\makebox(0,0){$\star$}}
\end{picture}
\begin{center}
{\bf Fig. 1b}
\end{center}

\vspace{1.3 cm}

\setlength{\unitlength}{0.240900pt}
\ifx\plotpoint\undefined\newsavebox{\plotpoint}\fi
\sbox{\plotpoint}{\rule[-0.200pt]{0.400pt}{0.400pt}}%
\begin{picture}(1500,629)(0,0)
\font\gnuplot=cmr10 at 10pt
\gnuplot
\sbox{\plotpoint}{\rule[-0.200pt]{0.400pt}{0.400pt}}%
\put(220.0,143.0){\rule[-0.200pt]{4.818pt}{0.400pt}}
\put(198,143){\makebox(0,0)[r]{2}}
\put(1416.0,143.0){\rule[-0.200pt]{4.818pt}{0.400pt}}
\put(220.0,203.0){\rule[-0.200pt]{4.818pt}{0.400pt}}
\put(198,203){\makebox(0,0)[r]{3}}
\put(1416.0,203.0){\rule[-0.200pt]{4.818pt}{0.400pt}}
\put(220.0,262.0){\rule[-0.200pt]{4.818pt}{0.400pt}}
\put(198,262){\makebox(0,0)[r]{4}}
\put(1416.0,262.0){\rule[-0.200pt]{4.818pt}{0.400pt}}
\put(220.0,322.0){\rule[-0.200pt]{4.818pt}{0.400pt}}
\put(198,322){\makebox(0,0)[r]{5}}
\put(1416.0,322.0){\rule[-0.200pt]{4.818pt}{0.400pt}}
\put(220.0,382.0){\rule[-0.200pt]{4.818pt}{0.400pt}}
\put(198,382){\makebox(0,0)[r]{6}}
\put(1416.0,382.0){\rule[-0.200pt]{4.818pt}{0.400pt}}
\put(220.0,442.0){\rule[-0.200pt]{4.818pt}{0.400pt}}
\put(198,442){\makebox(0,0)[r]{7}}
\put(1416.0,442.0){\rule[-0.200pt]{4.818pt}{0.400pt}}
\put(220.0,501.0){\rule[-0.200pt]{4.818pt}{0.400pt}}
\put(198,501){\makebox(0,0)[r]{8}}
\put(1416.0,501.0){\rule[-0.200pt]{4.818pt}{0.400pt}}
\put(220.0,561.0){\rule[-0.200pt]{4.818pt}{0.400pt}}
\put(198,561){\makebox(0,0)[r]{9}}
\put(1416.0,561.0){\rule[-0.200pt]{4.818pt}{0.400pt}}
\put(220.0,113.0){\rule[-0.200pt]{0.400pt}{4.818pt}}
\put(220,68){\makebox(0,0){0.93}}
\put(220.0,541.0){\rule[-0.200pt]{0.400pt}{4.818pt}}
\put(342.0,113.0){\rule[-0.200pt]{0.400pt}{4.818pt}}
\put(342,68){\makebox(0,0){0.94}}
\put(342.0,541.0){\rule[-0.200pt]{0.400pt}{4.818pt}}
\put(463.0,113.0){\rule[-0.200pt]{0.400pt}{4.818pt}}
\put(463,68){\makebox(0,0){0.95}}
\put(463.0,541.0){\rule[-0.200pt]{0.400pt}{4.818pt}}
\put(585.0,113.0){\rule[-0.200pt]{0.400pt}{4.818pt}}
\put(585,68){\makebox(0,0){0.96}}
\put(585.0,541.0){\rule[-0.200pt]{0.400pt}{4.818pt}}
\put(706.0,113.0){\rule[-0.200pt]{0.400pt}{4.818pt}}
\put(706,68){\makebox(0,0){0.97}}
\put(706.0,541.0){\rule[-0.200pt]{0.400pt}{4.818pt}}
\put(828.0,113.0){\rule[-0.200pt]{0.400pt}{4.818pt}}
\put(828,68){\makebox(0,0){0.98}}
\put(828.0,541.0){\rule[-0.200pt]{0.400pt}{4.818pt}}
\put(950.0,113.0){\rule[-0.200pt]{0.400pt}{4.818pt}}
\put(950,68){\makebox(0,0){0.99}}
\put(950.0,541.0){\rule[-0.200pt]{0.400pt}{4.818pt}}
\put(1071.0,113.0){\rule[-0.200pt]{0.400pt}{4.818pt}}
\put(1071,68){\makebox(0,0){1}}
\put(1071.0,541.0){\rule[-0.200pt]{0.400pt}{4.818pt}}
\put(1193.0,113.0){\rule[-0.200pt]{0.400pt}{4.818pt}}
\put(1193,68){\makebox(0,0){1.01}}
\put(1193.0,541.0){\rule[-0.200pt]{0.400pt}{4.818pt}}
\put(1314.0,113.0){\rule[-0.200pt]{0.400pt}{4.818pt}}
\put(1314,68){\makebox(0,0){1.02}}
\put(1314.0,541.0){\rule[-0.200pt]{0.400pt}{4.818pt}}
\put(1436.0,113.0){\rule[-0.200pt]{0.400pt}{4.818pt}}
\put(1436,68){\makebox(0,0){1.03}}
\put(1436.0,541.0){\rule[-0.200pt]{0.400pt}{4.818pt}}
\put(220.0,113.0){\rule[-0.200pt]{292.934pt}{0.400pt}}
\put(1436.0,113.0){\rule[-0.200pt]{0.400pt}{107.923pt}}
\put(220.0,561.0){\rule[-0.200pt]{292.934pt}{0.400pt}}
\put(45,337){\makebox(0,0){\mbox{\boldmath{$h_{t}^0h_b^0$}}}}
\put(828,23){\makebox(0,0){\mbox{\boldmath{$h_{t}h_b$}}}}
\put(828,606){\makebox(0,0){{ $h_th_b$- fixed point
structure for \mbox{\boldmath{$h_{t}^0/h_{b}^0=1.2,1.001$}}}}}
\put(220.0,113.0){\rule[-0.200pt]{0.400pt}{107.923pt}}
\put(589,135){\makebox(0,0){$+$}}
\put(628,143){\makebox(0,0){$+$}}
\put(662,151){\makebox(0,0){$+$}}
\put(694,159){\makebox(0,0){$+$}}
\put(724,167){\makebox(0,0){$+$}}
\put(751,176){\makebox(0,0){$+$}}
\put(775,185){\makebox(0,0){$+$}}
\put(798,194){\makebox(0,0){$+$}}
\put(820,203){\makebox(0,0){$+$}}
\put(839,213){\makebox(0,0){$+$}}
\put(858,223){\makebox(0,0){$+$}}
\put(875,233){\makebox(0,0){$+$}}
\put(891,243){\makebox(0,0){$+$}}
\put(905,253){\makebox(0,0){$+$}}
\put(919,264){\makebox(0,0){$+$}}
\put(932,275){\makebox(0,0){$+$}}
\put(944,287){\makebox(0,0){$+$}}
\put(956,298){\makebox(0,0){$+$}}
\put(967,310){\makebox(0,0){$+$}}
\put(977,322){\makebox(0,0){$+$}}
\put(986,335){\makebox(0,0){$+$}}
\put(995,347){\makebox(0,0){$+$}}
\put(1004,360){\makebox(0,0){$+$}}
\put(1012,373){\makebox(0,0){$+$}}
\put(1019,386){\makebox(0,0){$+$}}
\put(1027,400){\makebox(0,0){$+$}}
\put(1033,414){\makebox(0,0){$+$}}
\put(1040,428){\makebox(0,0){$+$}}
\put(1046,442){\makebox(0,0){$+$}}
\put(1052,457){\makebox(0,0){$+$}}
\put(1057,471){\makebox(0,0){$+$}}
\put(704,158){\makebox(0,0){$\star$}}
\put(736,167){\makebox(0,0){$\star$}}
\put(764,176){\makebox(0,0){$\star$}}
\put(791,186){\makebox(0,0){$\star$}}
\put(815,196){\makebox(0,0){$\star$}}
\put(837,206){\makebox(0,0){$\star$}}
\put(858,217){\makebox(0,0){$\star$}}
\put(877,228){\makebox(0,0){$\star$}}
\put(894,239){\makebox(0,0){$\star$}}
\put(910,250){\makebox(0,0){$\star$}}
\put(926,262){\makebox(0,0){$\star$}}
\put(940,274){\makebox(0,0){$\star$}}
\put(953,287){\makebox(0,0){$\star$}}
\put(965,299){\makebox(0,0){$\star$}}
\put(976,312){\makebox(0,0){$\star$}}
\put(987,325){\makebox(0,0){$\star$}}
\put(997,339){\makebox(0,0){$\star$}}
\put(1006,353){\makebox(0,0){$\star$}}
\put(1015,367){\makebox(0,0){$\star$}}
\put(1023,382){\makebox(0,0){$\star$}}
\put(1031,396){\makebox(0,0){$\star$}}
\put(1038,411){\makebox(0,0){$\star$}}
\put(1045,427){\makebox(0,0){$\star$}}
\put(1052,442){\makebox(0,0){$\star$}}
\put(1058,458){\makebox(0,0){$\star$}}
\put(1064,475){\makebox(0,0){$\star$}}
\put(1069,491){\makebox(0,0){$\star$}}
\put(1075,508){\makebox(0,0){$\star$}}
\put(1080,525){\makebox(0,0){$\star$}}
\put(1084,543){\makebox(0,0){$\star$}}
\put(1089,560){\makebox(0,0){$\star$}}
\end{picture}
\begin{center}
{\bf Fig. 1c}
\end{center}

\section{Analytic Solutions for $A_{t}$ and $A_{b}$}

The differential equations which govern the evolution of the
soft scalar masses of the third generation and the two Higgs
 mass parameters $m_{H_{1,2}}$, are well known.
In order to solve them, we need first a solution for
the $A_{t,b,\tau}$ trilinear mass terms. Ignoring the
$\tau$--Yukawa
coupling, we can write  the $A_t$ and $A_b$  evolution equations
as follows
\begin{eqnarray}
\frac{d A_t}{d t} & = &
\frac{1}{8 \pi^2} \left (
6 h_{t}^2 A_t + h_{b}^2 A_b + \hat G_{Q} m_{1/2} \right )
\nonumber \\
\frac{d A_b}{d t} & = &
\frac{1}{8 \pi^2} \left (
 6 h_{b}^2 A_b + h_{t}^2 A_t  + \hat G_{B} m_{1/2}
\right )
\end{eqnarray}
where $\hat G_{Q}$ and $\hat G_{D}$ are given by
\beq
\hat G_Q = \frac{4 \pi}{a_{G}}
\sum^{3}_{i=1}\hat c_{Q}^{i} a_{i}^2\,\, ,\qquad
\hat G_B = \frac{4 \pi}{a_{G}}
\sum^{3}_{i=1}\hat c_{B}^{i} a_{i}^2
\eeq
where we have taken into account that the gluino masses
are given by
\beq
M_{i} \approx m_{1/2} \frac{a_{i}}{a_{G}}\label{glu}
\eeq
$a_{G}$ being the common coupling at unification scale.
Since the small corrections that arise from the $U(1)$ factors
may be ignored we have $\hat G_{Q} \approx \hat G_{D}$.

To solve this system we follow the lines of \cite{fl1},
where the system ${\cal M}_U^2 - {\cal M}_D^2 $
(ignoring $A_t$, $A_b$) which has a similar structure has
 been solved \footnote{
We come back to the system
${\cal M}_U^2 - {\cal M}_D^2 $
in section 4.}. We initially separate the running of the
Yukawa couplings by rewriting $A_{t}$ and $A_{b}$ as
\beq
A_t  = \tau X_t, \; \; \;
A_b = \sigma X_b ,
\label{eq:30ab}
\eeq
where
\beq
 \tau = exp\Big\{\frac{3}{4\pi^2}
 \int^t_{t_0}  h_t^2\, dt^\prime\Big\} \label{tau},
\; \; \; \; \;
 \sigma = exp \Big\{\frac{3}{4\pi^2}
\int^t_{t_0}  h^2_b dt^\prime\Big\}
\label{sigma}
\eeq
Defining the $2\times 2$ matrix
\begin{eqnarray}
{\cal H}(t)
=\gamma_Q^2(\sigma \tau )^{\frac 7{12}}
\left[\begin{array}{cc}0&y_0(\frac{\sigma}{\tau})^{\frac{17}{12}} \\
x_0(\frac{\tau}{\sigma})^{\frac{17}{12}} &0
\end{array}
 \right] \label{eq:3c}
\end{eqnarray}
and the function
\beq
h(u)=\left(\frac{x_0}{y_0}\right)^{\frac 65}
\left(\frac{\sqrt{1+u}-1}{\sqrt{1+u}+1}\right)^{\frac{17}{10}}
\eeq
the $A_t-A_b$ system can be written as
\begin{eqnarray}
 \frac{d}{du}
\left(\begin{array}{c} X_t \\ X_b \end{array}
 \right) = -\frac 1{10}\frac 1{\sqrt{u^2+u}}
\left[\begin{array}{cc}0&h(u) \\
h(u)^{-1} &0
\end{array}
 \right]
 \left(\begin{array}{c} X_t \\ X_b
\end{array} \right) +
\frac {m_{1/2}}{8\pi^2} \frac {dt}{du}\left(\begin{array}{c}
\frac{\hat G_Q}{\tau} \\ \frac{\hat G_B}{\sigma} \end{array}
\right)
\label{eq:32a}
\end{eqnarray}
Note that $h(u) $  in the  large $tan\beta$ case , (i.e.
 $h_{t,0}\sim h_{b,0}$) is approximately
constant in most of the range of integration.

Then, in the case $u \gg 1$ we can write
an approximate analytic solution  as follows
\begin{eqnarray}
A_t & \approx & \frac{\tau}{2}
\left\{
(A_t^0+h_0 A_b^0)\rho +
(A_t^0-h_0 A_b^0)\frac{1}{\rho}\right\}-m_{1/2}<J_{\tau}>
\label{At}\\
A_b
&\approx
&\frac{\sigma}{2 h_0} \left\{(A_t^0+h_0A_b^0)\rho -
(A_t^0-h_0A_b^0)\frac{1}{\rho}\right\} - m_{1/2}<J_{\sigma}>
\label{Ab}
\end{eqnarray}
with
$$\rho = (\frac {tan\phi}{tan\phi_0})^{\frac 15}
\qquad     sin 2\phi = ({1+u})^{-\frac 12}$$
while $ h_0 = h(u\ra \infty )$.

{}Furthermore, in the limit $u \gg 1$, the  integrals
 $<J_{\tau ,\sigma}>$ are given by
\beq
<J_{\tau}> = \tau \int_t^{t_0} \frac {\hat G_Q({s})}
{\tau{({s})}} \frac 1{\rho (s)}
d{s} \label{Itau2}
\eeq
and similarly for $J_{\sigma}$.
A simple inspection of the above formulae shows that for
reasonable initial $A_{t,b}$ values the terms proportional
to $m_{1/2}$ dominate. Thus, to a good approximation we may
write $A_{t}\approx -m_{1/2}<J_{\tau}>$
and $ A_{b}\approx -m_{1/2}<J_{\sigma}>$.

The semi-analytic expressions (\ref{At}-\ref{Ab}) are going
to be used in  the following sections, in order to compute
the contributions to sparticle masses, as well as
the corrections to the bottom mass from superparticle
contributions.

\section{Predictions for sparticle masses and comparison
with the exact solutions of the RGE}

Having obtained the solutions for $A_t$ and $A_b$ (thus for
${\cal M}_U^2$ and ${\cal M}_D^2$) we may
calculate contributions to superparticle masses, by solving
the system
\begin{eqnarray}
{\cal M}_U^2\equiv \tilde m_{Q_L}^2+
                          \tilde m_U^2+m_{H_2}^2+A_t^2 \nonumber \\
{\cal M}_D^2\equiv \tilde m_{Q_L}^2+
                          \tilde m_D^2+m_{H_1}^2+A_b^2
\end{eqnarray}
with initial conditions
 ${\cal M}_{(U,D)_0}^2=
\xi_{(U,D)} m_0^2\,$, where
$m_0$ is a common scalar mass at the unification scale,
$\xi_U\equiv \xi_{H_2} +\xi_Q + \xi_{t^c}$ and
$\xi_D\equiv \xi_{H_1} +\xi_Q + \xi_{b^c}$
Taking into account the renormalisation
group equations for the squark and Higgs fields\footnote{
In the renormalisation group equations of the squark
and Higgs fields, for non-universal initial conditions,
contributions proportional to
$\frac{\alpha _1}{2\pi }S$
where $S(t)=\frac{\alpha_1(t)}{\alpha_{1,0}}Tr\left[Ym^2\right]$
are obtained.
However, these contributions cancel in the
equations that describe the sums
${\cal M}_U^2$ and ${\cal M}_D^2$,
due to the invariance of the $U(1)$ Yukawa Lagrangian.
The non-universal initial conditions are still manifest,
through the factors $\xi_{i}$
.}
, we obtain
\begin{eqnarray}
 \frac{d{\cal M}_U^2}{dt}  &=&
\frac{1}{8\pi^2}
 \Big\{6{\cal M}_U^2 h^2_t +{\cal M}_D^2h^2_b - G_U^0 m_{1/2}^2
\Big \}
+ \frac{d A_t^2}{d t}
\label{eq:28} \\
\frac{d{\cal M}_D^2}{dt}  &=& \frac{1}{8\pi^2}
\Big\{{\cal M}_U^2h^2_t + 6{\cal M}_D^2h^2_b - G_D^0  m_{1/2}^2
\Big\}
+ \frac{d A_b^2}{d t}
 \label{eq:29}
\end{eqnarray}
where $G_U^0 = G_Q + G_{H_2} +
G_{U^c}$ and $G_D^0 = G_Q + G_{H_1} + G_{B^c}$.
Taking into account the  $A_t$ and $A_b$ contributions,
the solution of the system to first order reads
\begin{eqnarray}
\frac{{\cal M}_U^2}{m_0^2}
&\approx
&\frac{\tau}{2} \left\{(\xi_{U}+h_0\xi_{D})\rho +
(\xi_{U}-h_0\xi_{D})\frac{1}{\rho}\right\}+\xi_{1/2}<I_{\tau}>
-<A_{\tau}> \\
\frac{{\cal M}_D^2}{m_0^2}
&\approx
&\frac{\sigma}{2 h_0} \left\{(\xi_{U}+h_0\xi_{D})\rho -
(\xi_{U}-h_0\xi_{D})\frac{1}{\rho}\right\} + \xi_{1/2}<I_{\sigma}>
- <A_{\sigma}>
\end{eqnarray}
where $\xi_U\equiv \xi_{H_2} +\xi_Q + \xi_{t^c}\,\,$,
$\xi_D\equiv \xi_{H_1} +\xi_Q + \xi_{b^c}$.
$<I_{\tau}>$ is given by
\beq
<I_{\tau}>=\tau \int_t^{t_0}\frac {G_U({t^\prime})}{\tau{({t^\prime})}}
\frac 1{\rho (t^\prime)}
d{t^\prime} . \label{Itau}
\eeq
A similar expression is obtained
for $<I_{\sigma}>$ with the replacements $\tau\ra \sigma $
and $G_U \ra G_D$.  Finally,
\beq
<A_{\tau}> =
2 \tau \int_{t}^{t_{0}} \left (
 \frac{A_t(s)} {\tau(s)}\frac{A'_t(s)}
{\rho(s)} \right )
ds
\eeq
Here, $A_t$ is approximately given by
\beq
A_t(t) = -m_{1/2} \tau(t) \int_{t}^{t_{0}} \left (
 \frac{\hat G_Q(s)} {\tau(s)}\frac{1}
{\rho(s)} \right )
ds
\eeq
and
\beq
A'_t(t) = -m_{1/2} \tau'(t) \int_{t}^{t_{0}}
 \frac{\hat G_Q(s)} {\tau(s)}\frac{1}
{\rho(s)} + m_{1/2} \tau (t)
\frac{\hat G_Q(t)} {\tau(t)}\frac{1}
{\rho(t)}
\eeq
where primes denote derivatives with respect to the logarithm of
the scale $t = log(\mu )$.
Similarly, $<A_{\sigma}>$ is found by the proper substitutions.

Integrating the sums, we obtain
\begin{eqnarray}
{\cal M}_U^2 - {\cal M}_{U,0}^2 - C_U(t)m_{1/2}^2 = - 6 J_U - J_D\label{al1}
+ A_t^2 - A_{t,0}^2 \\
{\cal M}_D^2 - {\cal M}_{D,0}^2 - C_D(t)m_{1/2}^2 =  - J_U - 6 J_D\label{al2}
+ A_b^2 - A_{b,0}^2
\end{eqnarray}
with $J_I(t) = \int h_I^2 {\cal M}_I^2 dt$, $I=U,D$. Now, the unknown integrals
 $J_I(t)$ can be expressed in terms of the already calculated functions, their
initial conditions and known gauge functions, from the simple algebraic system
(\ref{al1},\ref{al2}).
{}As an example, for the up--squark running mass squared  we have
\beq
\tilde{m}_{Q}^2 = (\xi_{Q} + C_Q(t)\xi_{1/2})
 m_0^2 - J_D(t) - J_U(t) + I'_S\label{hig2}
\eeq
with $I'_S$ representing the integral of the $S$--contribution in the
case of non -- universality and
\begin{equation}
C_Q(t)=\sum^3_{i=1}
\frac{c_i^Q}{2b_i\alpha^2_{i_G}}
\left(\alpha_i^2(t_1)-\alpha_i^2(t)\right)
\end{equation}

In  {\it table 1} we compare the superparticle masses  obtained
from the analytic formulae, with those when solving numerically the
renormalisation group equations: In the first column, we give mass
terms, for a numerical solution of the renormalisation group equations,
and at a scale $\approx 250$ GeV, when the heavier superparticle,
which in our case is the gluino, decouples from the spectrum.
The initial conditions we take are
$$A_t^0 = A_b^0=1 ,$$
$$h_{t}^0 = 2.0,\;\; h_{t}^0 / h_{b}^0 \;\; \approx 1.1 ,$$
$$\xi_{H_1}=4.0, \;\; \xi_{H_2} = 1.0 ,$$
$$\xi_{d^c}=1.0,\;\; \xi_{u^c}=1.0,\;\; \xi_{q}=1.0 ,$$
and ignore for simplicity the $I_S$ contribution to scalar masses
(of course we do the same in the analytical solutions).
In the second column appear the analytic solutions,
when we include the $A_t, A_b$ contributions, while
in the third the solutions when these contributions are neglected.
{}Finally, for comparison, in the fourth column we give
the superparticle masses that are found numerically by the
renormalisation group equations, when we ignore the
$A_{t,b}$ contributions. Here we have taken
$m_0 = m_{1/2} $ = 100 GeV
and the superparticle masses are given in GeV.
The effect of the $\tau$ coupling has been ignored in
the numerical solutions of the renormalisation group equations
as well, for a better comparison.
Comparing the analytic with the numerical solutions,
we find that the total relative error is at most $2-3\%$.
Part of this small error in the case where we include the
$A_{t,b}$ contributions arises because we wanted to keep the
expressions as simple as possible and therefore have ignored
the contribution of the homogeneous part of the
$A_{t,b}$ solutions in the superparticle
masses (but not in the calculation of the $A$'s themselves).
Had we included
the effect of the $\tau$ Yukawa coupling in the
numerical solution of the
renormalisation group equations, the shift to the results
is at most $10\%$. These observations
are in agreement with the ones obtained in the Appendix A
of \cite{kpz2}.
\begin{table}
\begin{center}
\begin{tabular}
{|c|cccc|}
\hline
$m_{gluino}$ &  252 & 252 & 252 & 252 \\ \hline
$m_{Q_L}$ & 182 & 185 & 189 & 188 \\
$m_{t_R}$ & 185 & 188 & 193 & 191 \\
$m_{b_R}$ & 153 & 157 & 162 & 160 \\
$A_t$ & 143 & 147 & -- & -- \\  \hline
$A_{b}$ & 144 & 146 & -- & -- \\
\hline
\end{tabular}
\end{center}

{{\bf Table 1 : } {\small Some sparticle masses (in GeV) obtained
using analytic (second column) or numerical (third column)
solutions. The last two columns refer to the corresponding
cases but when the $A_{b,t}$ terms are ignored.}}
\label{table:I}
\end{table}

Let us finally make the following observation:
In the equation for the product of the Yukawa couplings,
we see that the larger the Yukawas and
the smaller the difference between $h_{t}-h_{b}$, the stronger the
fixed point behaviour becomes. Moreover, in the equations for the
scalar masses, we observe that the
smaller the difference between $h_{t}^0-h_{b}^0$, the less important
the contribution of the homogeneous part of the solution
(which includes the main dependence from initial
conditions for the Yukawa couplings, as
the integrals $<I_{\tau,\sigma}>$ undergo a
smaller change) becomes. This indicates that for low energy
Yukawas which correspond to $h_t^0 = h_b^0$ and values
close to the non-perturbative region for the couplings, one in
principle expects to obtain the minimal variations of the effective
potential under small displacements of the couplings.
In \cite{kpz2} it has been shown that
the effective potential exhibits a minimum related to the
infrared fixed line, while for non--universal boundary
conditions for the scalars, minima which correspond to
$h_t=h_b$ will also be favored. This we think can be understood
by looking at the regions where the expressions  we
have derived exhibit the strongest attraction to the
 fixed points.

\section{Corrections to the bottom mass}

It has been found that in the large tan$\beta$ regime, there
are large contributions to the running bottom
quark  mass $m_b$.
These are given by \cite{bl,sc}
\begin{eqnarray}
m_b = h_b v_1 \left( 1 + \delta m_b \right).
\end{eqnarray}
where  $\delta m_b =\epsilon_b  tan\beta$ and
\begin{eqnarray}
\epsilon_b  = \frac{\mu}{16\pi^2}  \left(
 \frac{8}{3}g^2_3 M_{\tilde{g}}
 I(m_{\tilde{b},1}^2, m_{\tilde{b},2}^2,
M_{\tilde{g}}^2)  +  { h^2_t } A_t
I(m_{\tilde{t},1}^2, m_{\tilde{t},2}^2,\mu^2)\right),
\end{eqnarray}
and the integral function I(a,b,c) is given by
\begin{equation}
I(a,b,c) = \frac{a b \ln(a/b) + b c \ln(b/c) + a c \ln(c/a) }
{( a - b )( b - c )( a - c )},
\end{equation}
with
$M_{\tilde{g}}$ and $m_{\tilde{b},i}$  ($m_{\tilde{t},i}$)
being the gluino and sbottom (stop) eigenstate masses respectively.

To calculate these corrections we substitute the
relevant expressions in the above formulae.
In \cite{fl1}, expressions for the scalar masses have been derived
(and these expressions are not sensitive to $A_{t}$, $A_{b}$).
Here, we have derived an equation for $A_t$, while
the gluino mass is approximately given by (\ref{glu}).
For the low energy value of $\mu$ we obtain the analytic expression
\beq
\mu^2 \approx - m_{H_2}^2-\frac 12 m_Z^2
\label{eqmu}
\eeq
while the renormalisation group equation can be integrated to give\barr
\mu &=&\mu_G (\frac{u_0}{u})^{\frac 3{10}}
 \prod_{j=1}^3 \left(\frac{\alpha_{j,0}}
 {\alpha_{j}}\right)^{c_\mu^j/2b_j},
\err
with $\{c_\mu^i \}_{i=1,2,3} = \left\{ \frac{3}{5},3,0\right\}$.

Then, we find the magnitude of the bottom corrections,
for solutions at the fixed point ($h_G=2.0$), as well as
for small deviations from the fixed point ($h_{G}=1.0$),
while we keep $h_{t}^0/h_b^0 \approx 1.1$.
The relevant quantities appear in {\it tables
2 } and \it 3}. The masses in the tables are given in GeV.

As we increase the supersymmetry breaking scale,
$tan\beta$ slightly increases, in order to get the same low energy
 parameters.
(At the same time the unification scale drops slightly, while
the inverse gauge coupling at the unification scale increases,
by a small amount).

\begin{table}
\centering
\begin{tabular}
{|c|c|c|c|c|c|c|}
\hline
$tan\beta$ & $m_0=m_{1/2}$
& $\mu$ & $A_{t}$ & $I_1 (10^{-6})$ & $I_2(10^{-6})$
& $\delta m_b$
\\ \hline \hline
58.1 & 100 & 124 & -147 & 12.8 & 18.1 & 0.40 \\
59.3  & 150 & 192 & -208 & 6.2 & 8.5 & 0.41 \\
60.0 & 200 & 253 & -264 & 3.8 & 5.1  & 0.42 \\
60.3 & 250 & 310 & -316 & 2.6 & 3.4 & 0.43 \\
60.6 & 300 & 365 & -364 & 1.9 & 2.5 & 0.43 \\
\hline
\hline
\end{tabular}
\begin{center}
{{\bf Table 2 :} {\small Bottom mass corrections for $h_G\sim 2.0$.}}
\end{center}
\label{table:2}
\end{table}

\begin{table}
\centering
\begin{tabular}
{|c|c|c|c|c|c|c|}
\hline
$tan\beta$ & $m_0=m_{1/2}$
& $\mu$ & $A_{t}$ & $I_1 (10^{-6})$ & $I_2(10^{-6})$
& $\delta m_b$
\\ \hline \hline
55.5 & 100 & 119 & -153 & 12.1 & 18.0 & 0.34 \\
56.6  & 150 & 184 & -216 & 5.9 & 8.4 & 0.35 \\
57.1 & 200 & 243 & -274 & 3.5 & 5.0  & 0.36 \\
57.4 & 250 & 297 & -328 & 2.4 & 3.4 & 0.37 \\
57.5 & 300 & 348 & -377 & 1.8 & 2.5 & 0.37 \\
\hline
\hline
\end{tabular}
\begin{center}
{{\bf Table 3 : } {\small Bottom mass corrections for $h_G\sim 1.0$.}}
\end{center}
\label{table:3}
\end{table}

\section{Conclusions}

In this letter we have used simplified  analytic
solutions for the $h_t , h_b$ Yukawa
couplings in order to study the MSSM in the large
$tan\beta$ regime.
We have explored the regions of the parameter space
which lead to a fixed point structure and derived the evolution
of the Yukawas towards these fixed points. Using this information,
 one may identify the regions for the initial values of
the Yukawa couplings which lead to the strongest attraction
towards these infrared fixed points. Under these considerations,
top-bottom Yukawa coupling
equality, and values of the couplings  close to
the non-perturbative regime seem to be favoured.
Finally, we obtained corrections on the renormalised soft mass
terms due to the evolution of the trilinear  parameters
$A_{t}$ and $A_{b}$. Using these results, we
estimated analytically the  sparticle loop -- corrections
to the  bottom mass, which are important in the
large tan$\beta$ scenario.
In agreement with previous calculations we find that
the maximal corrections arise at the fixed point.

\vspace*{2cm}

{\it G.K.L.  would like to thank the group of Centre de Physique
Theorique de l'  Ecole Polytechnique for kind hospitality.}

\newpage




\begin{thebibliography}{99}

\bibitem{susy}
H.~P. Nilles, Phys. Rep. {\bf 110}(1984)1;
\\
G.~G.~Ross, {\it Grand Unified Theories}, Benjamin Cummings
(1985);
\\
H.~E. Haber and G. ~L. Kane, Phys. Rep. {\bf 117}(1985)75;
\\
A. ~B. Lahanas and D. ~V. Nanopoulos, Phys. Rep. {\bf 145}(1987)1;
\\
S. Ferrara, ed., ``Supersymmetry'' (North-Holland, Amsterdam, 1987);

\bibitem{unif}
S. Dimopoulos and H. Georgi, Nucl. Phys. {\bf B193} (1981)150;
\\
J. Ellis, S. Kelley and D.V. Nanopoulos, Phys. Lett. {\bf B249}
(1990)441; Phys. Lett. {\bf B260} (1991)
131;
U. Amaldi, W. de Boer and H. F\"urstenau, Phys. Lett. {\bf B260}
(1991)
447;P. Langacker and M. Luo, Phys.Rev. {\bf D44} (1991) 817;

\bibitem{radi}
 K. Inoue et al., Prog. Theor. Phys. {\bf 68} (1982) 927;
L.E. Ib\'a\~nez, Nucl.Phys. {\bf B218} (1983) 514;
L.E. Ib\'a\~nez and C. L\'opez, Phys. Lett. {\bf B126} (1983) 54;
Nucl.Phys. {\bf B233} (1984) 511;
L. Alvarez-Gaume, J. Polchinsky and M. Wise, Nucl.Phys. {\bf
B221}
(1983) 495; L.E Ib\'a\~nez, C. L\'opez and C. Mu\~noz, Nucl.
Phys. {\bf B256} (1985)
218.
\bibitem{ross}B. Pendleton and G. G. Ross, Phys. Lett.{\bf B}98
(1981)291;
\\
C. T. Hill, Phys. Rev. {\bf D}24(1981)291;
\\
M. Lanzagorta and G. G. Ross,  Phys. Lett.{\bf B 349} (1995)319.
\bibitem{fl} E. G. Floratos  and  G.K. Leontaris,
 Phys. Lett. {\bf B336}(1994)194;

\bibitem{kpz1}
C. Kounnas, I. Pavel and F. Zwirner,
 Phys. Lett.{\bf B 335} (1994) 403.

\bibitem{kpz2}
C. Kounnas, I. Pavel, G. Ridolfi and F. Zwirner,
Phys. Lett. {\bf B 354} (1995) 322.

\bibitem{fl1}
E. G. Floratos and G. K. Leontaris, hep-ph/9503455 , IOA-320-95,
to be published in Nucl. Phys. B.

\bibitem{bl} T. Banks, Nucl. Phys. {\bf B 303} (1988) 172;\\
G. K. Leontaris, Phys. Lett. {\bf B 236} (1989) 179.
\bibitem{sc}L. J. Hall, R. Rattazzi and U. Sarid, Phys. Rev. {\bf D
50}(1994)7048;
M. Carena, M. Olechowski, S. Pokorski, and C. E. M. Wagner,
Nucl.Phys.{\bf B 426} (1994) 269.

\bibitem{BD}
P. Binetruy and E. Dudas,  Nucl. Phys. {\bf B 442} (1995)21.

\end{thebibliography}
\end{document}